\documentclass[aps,prl,twocolumn]{revtex4}
\usepackage{amsfonts}
\usepackage{amsmath}
\usepackage{amssymb}
\usepackage{graphicx}
\usepackage{color}

\begin{document}

\title{Anomalous Diffusion in a Dynamical Optical Lattice}
\author{Wei Zheng}
\affiliation{T.C.M. Group, Cavendish Laboratory, J. J. Thomson Avenue, Cambridge CB3 0HE,
United Kingdom}
\author{Nigel R. Cooper}
\affiliation{T.C.M. Group, Cavendish Laboratory, J. J. Thomson Avenue, Cambridge CB3 0HE,
United Kingdom}

\begin{abstract}
Motivated by experimental progress in strongly coupled atom-photon systems
in optical cavities, we study theoretically the quantum dynamics of atoms
coupled to a one-dimensional dynamical optical lattice. The dynamical
lattice is chosen to have a period that is incommensurate with that of an
underlying static lattice, leading to a dynamical version of the Aubry-Andr{%
\' e} model which can cause localization of single-particle wavefunctions.
We show that atomic wavepackets in this dynamical lattice generically spread
via anomalous diffusion, which can be tuned between super-diffusive and
sub-diffusive regimes. This anomalous diffusion arises from an interplay
between quantum localization and quantum fluctuations of the cavity field.
\end{abstract}

\date{12 Sep, 2017}

\maketitle

One of the most interesting directions of research in coherent quantum
systems concerns the collective dynamics of coupled atom-photon ensembles.
Such situations arise for cold atomic gases in optical cavities\cite{rev2013}
or waveguides\cite{nano_cavity2013,pc_cavity2014}, where strong coupling
between atomic motion and a photon field can be achieved.
%, in which the quantum dynamics of atomic motion is tied to the quantum dynamics of the electromagnetic field.
%
Coupling cold atoms even to a single cavity mode can dramatically change the
steady state of the atomic gas\cite%
{Esslinger2010PT,Esslinger2012roton,Esslinger2015roton,Hemmerich2015BHM,Hemmerich2015CBHM,Esslinger2016CBHM,Ritsch2002,Domokos2010,Simons2009,Goldbart2010,Simons2014,Piazza2014,Zhai2014,Yi2014,Pu2014,Yu2015,Yi2015,hopping2015,Kollath2016,Zhai2015,hopping2016,CBHM2016}%
and lead to interesting nonequilibrium dynamics\cite%
{Esslinger2008Dyn,Hemmerich2015Dyn,Hemmerich2017Bloch,ComOLDyn2005,Keeling2010,DampingInCavity2014,Critical_Dyn2013,Prethermalization2014,LimitCycle2015,Our2016,HoppingDyn2016}%
.

A transversely pumped Bose-Einstein condensate in a single-mode cavity can
undergo a phase transition into a self-organized ``superradiant'' state, in
which the cavity mode becomes highly occupied and generates a cavity-induced
superlattice potential on the atoms. For current experiments\cite%
{Hemmerich2015CBHM,Esslinger2016CBHM} this dynamical cavity-induced
superlattice is commensurate with an underlying static optical lattice,
therefore giving rise to a supersolid phase with extended Bloch waves.
However, one can readily envisage situations in which the cavity-induced
superlattice is incommensurate with the underlying static lattice. This
leads to the interesting possibility that the cavity-induced superlattice
leads to {localization} of the single-particle states. Indeed, several
theoretical works have studied the steady state of cold atoms in such
settings \cite{Zhang2011,CavityBoseGlass2013,CAAM2016}, and have found a
self-organized localization-delocalization transition within a mean-field
approximation.

In this paper, we show that the motion of atoms in a cavity-induced
incommensurate lattice is {qualitatively} affected by the quantum
fluctuations of the cavity field, leading to long-time behaviour that is not
captured by mean-field theories. Specifically, we show that the atomic
motion exhibits anomalous diffusion, in which the width of the wavepacket $%
\sigma $ grows with time as $\sigma \sim t^{\gamma }$ with $0<\gamma <1$.
Anomalous diffusion exists widely in both classical and quantum systems. In
classical random walks, anomalous diffusion is mostly associated with the
failure of the central limit theorem and the presence of long-tailed
distributions\cite{CAD1999,CAD_Rev2000,CAD_Rev2005}.
%For example, in L\'{e}vy flights,
%a broad distribution of hopping distances results in a superdiffusive
%behaviour\cite{Levy_fly1939,Levy_fly_book2002}.
On the other hand, in closed quantum systems, anomalous diffusion is
typically connected to the multifractal nature of eigenstates\cite%
{QAD_QC1996,QAD_floquet2017}. Quantum anomalous diffusion is also predicted
close to many body localization\cite{QAD_MBL2015,QAD_MBL2017}. Due to the
cavity loss, our dynamical lattice is an open system, so this anomalous
diffusion cannot be explained by multifractal eigenstates. Instead, using
the quantum trajectory picture, we show that the dynamics can be viewed as a
form of L\'{e}vy walk with rests\cite{Levy_walk_Rev2015,Levy_walk_book2011}.
This explanation relies both on quantum fluctuations of the cavity field and
on quantum localization in the incommensurate potential, so is an inherently
quantum phenomenon. We predict that evidence of this anomalous transport can
be found in long-tailed distributions of photon correlations in the cavity
field.

\textit{Model}. We consider spinless atoms trapped by an optical lattice in
a high-Q cavity (Fig.~1), both aligned along the $x$-direction. Two
counterpropagating pump lasers are shone on the atom cloud from the $z$%
-direction. Denoting the cavity field operator by $\hat{a}$, the net
potential on the atoms is $V=A\cos ^{2}(k_{\mathrm{o}}x)+B\hat{a}^{\dag }%
\hat{a}\cos ^{2}(k_{\mathrm{c}}x+\phi )+C(\hat{a}+\hat{a}^{\dag })\cos (k_{%
\mathrm{p}}z)\cos (k_{\mathrm{c}}x+\phi )$. Here $A$ is proportional to the
optical lattice intensity, $B$ is the cavity-atom coupling strength ($\phi $
controls the relative positions of the optical lattice and the cavity mode),
and $C$ is the pump-cavity coupling proportional to the amplitude of the
pump laser. We consider the transverse confinement to be sufficiently large
that the transverse motion is frozen out. For deep enough lattices, we
obtain a one dimensional tight-binding model as
\begin{eqnarray}
H &=&\Delta \hat{a}^{\dag }\hat{a}-J\sum_{j=1}^{L}\left( \hat{c}_{j+1}^{\dag
}\hat{c}_{j}+h.c.\right)   \notag \\
&&+\lambda \left( \hat{a}+\hat{a}^{\dag }\right) \sum_{j=1}^{L}u_{j}\hat{c}%
_{j}^{\dag }\hat{c}_{j}+U\hat{a}^{\dag }\hat{a}\sum_{j=1}^{L}u_{j}^{2}\hat{c}%
_{j}^{\dag }\hat{c}_{j},  \label{Ham}
\end{eqnarray}%
where $\hat{c}_{j}^{(\dag )}$ are the atomic field operators on lattice site
$j$; $\Delta =\omega _{\mathrm{c}}-\omega _{\mathrm{p}}$ is the detuning of
the cavity mode; $u_{j}=\cos (2\pi \beta j+\phi )$, $\beta =k_{\mathrm{c}%
}/2k_{\mathrm{o}}$; $U$ and $\lambda$ are the projections of $B$ and $C$
onto the Wannier functions. We have ignored interactions between atoms, as
can be realized by a Feshbach resonance for bosons or for spinless fermions
with contact interactions. Due to the leaking of photons from the cavity,
the system should be described by the quantum master equation
\begin{equation}
\partial _{t}\rho =-i\left[ H,\rho \right] +\kappa \left( 2\hat{a}\rho \hat{a%
}^{\dag }-\hat{a}^{\dag }\hat{a}\rho -\rho \hat{a}^{\dag }\hat{a}\right) ,
\label{QME}
\end{equation}%
where $2\kappa $ is the loss rate of the cavity photons.

If the cavity were directly driven by another pump laser, such that the
cavity field is a coherent state $\hat{a}\rightarrow \alpha $, then the
particles would experience a static effective potential $V_{\mathrm{eff}%
}(\alpha )=\sum_{j=1}^{L}\left[ 2\lambda \mathrm{Re}(\alpha
)u_{j}+U\left\vert \alpha \right\vert ^{2}u_{j}^{2}\right] \hat{c}_{j}^{\dag
}\hat{c}_{j}$. In the case where $\beta $ is a irrational number and $U=0$,
this reproduces the celebrated Aubry-Andr\'{e} model, which exhibits a
delocalization-localization transition for all the eigenstates\cite{AA1980}.
Even when $U\neq 0$, this transition still survives, but now with mobility
edges in the energy spectrum\cite{Dassarma2010}. We are interested in cases
without this direct drive, in which the cavity has its own quantum dynamics
and the atoms feel a dynamical potential.

\begin{figure}[tbp]
\includegraphics[width=2.8in]
{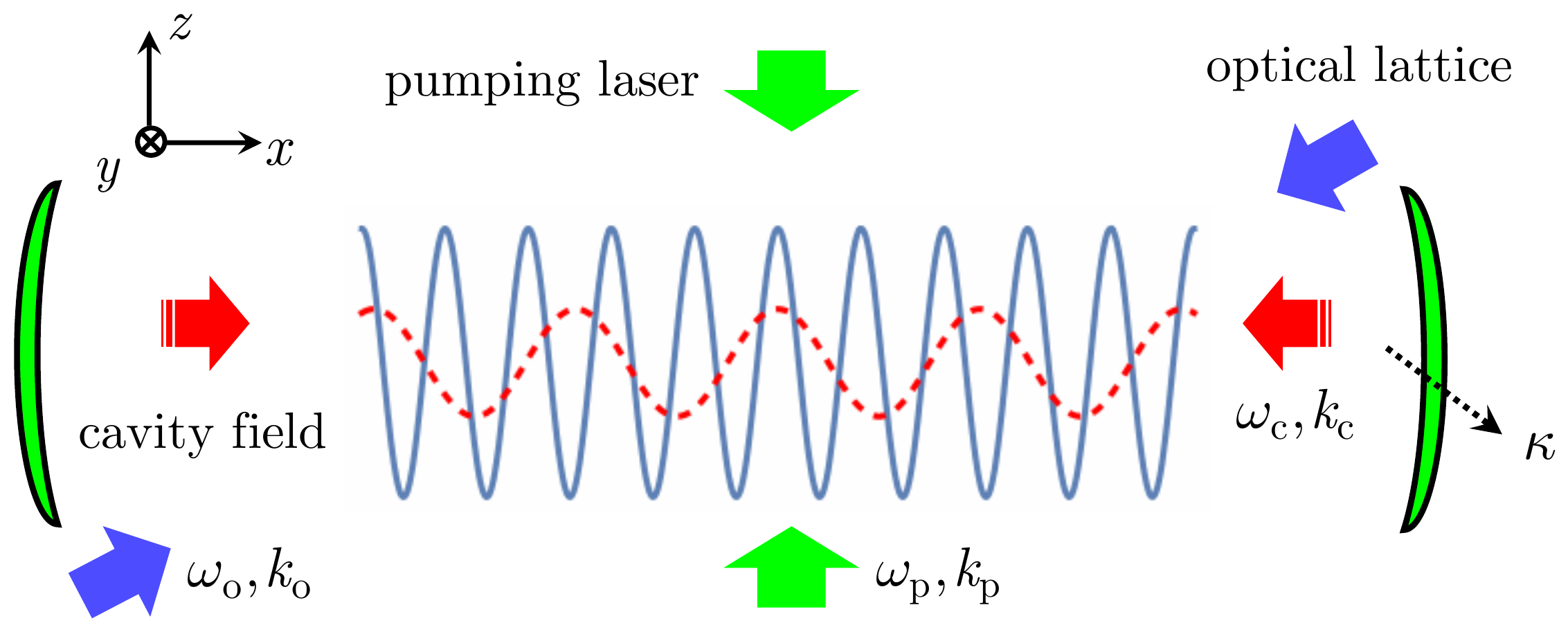}
\caption{Schematic diagram of the experimental setup. Atoms in a optical
lattice and a standing wave cavity is driven by a transverse laser. The
frequency $\omega _\mathrm{p}$ of the pump laser is far detuned from the
atomic transition line but close to the cavity mode frequency $\omega _\mathrm{c}$. }
\label{fig1}
\end{figure}

\textit{Mean field steady state}. From Eq. (\ref{QME}), one finds that the
mean cavity field $\alpha (t)=\left\langle \hat{a}(t)\right\rangle $ evolves
as $i\partial _{t}\alpha =\left( \Delta -i\kappa +UR\right) \alpha +\lambda
\Theta $, where $\Theta =\sum_{j}u_{j}\left\langle \hat{c}_{j}^{\dag }\hat{c}%
_{j}\right\rangle $, and $R=\sum_{j}u_{j}^{2}\left\langle \hat{c}_{j}^{\dag }%
\hat{c}_{j}\right\rangle $. We seek a steady state in which $\partial
_{t}\alpha =0$ and find $\alpha =-\frac{\lambda \Theta }{\Delta -i\kappa +UR}
$. The expectation value $\left\langle \hat{c}_{j}^{\dag }\hat{c}%
_{j}\right\rangle $ can be obtained from the ground state of the mean field
Hamiltonian in which the atoms experience the effective potential $V_{%
\mathrm{eff}}(\alpha )$.

\begin{figure}[tbp]
\includegraphics[width=3in]
{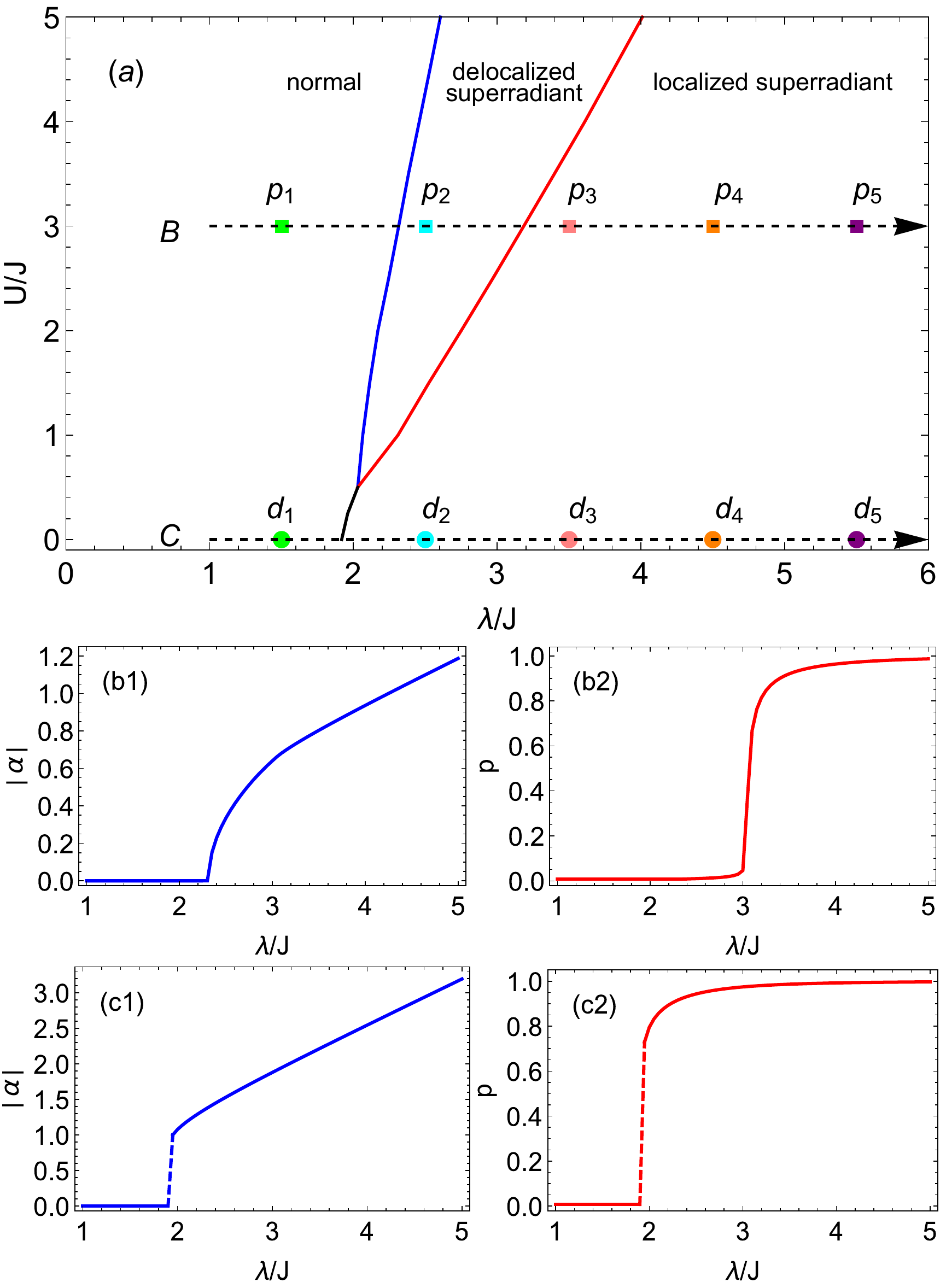}
\caption{(a) Phase diagram of mean field steady state. (b1), (b2) Second
order phase transition along the line B in (a). (b1) Mean cavity field $%
\protect\alpha $. (b2) Inverse partition ratio in real space $p$. (c1), (c2)
First order phase transition along the line C in (a). Here $L=201$, $\protect%
\beta =\left( \protect\sqrt{5}-1\right) /2$, $\protect\phi =\protect\pi /2$,
$\Delta /J=1$, and $\protect\kappa /J=1.2$. The square dots, $p_{i}$ ($%
i=1,\cdots 5$), represent the pumping strengths $\protect\lambda %
/J=1.5,2.5,3.5,4.5,5.5,$ respectively, and $U/J=3$; while the circle dots $%
d_{i}$ have the same pumping strengths but with $U/J=0$. }
\label{fig2}
\end{figure}

We consider one atom in the cavity, and numerically obtain the steady state
phase diagram, see Fig. \ref{fig2}. To describe the localization of the
particle, we calculate the inverse participation ratio in real space of the
atomic wavefunction, $p=\sum_{j=1}^{L}\left\vert \left\langle \hat{c}%
_{j}^{\dag }\hat{c}_{j}\right\rangle \right\vert ^{2}$. One notes that
highly localized density gives $p\sim 1$; while an extended wave function
has $p\sim 1/L$.

In the weak pumping regime, the system is in the \textquotedblleft normal"
phase, which has no superradiance, $\alpha =0$, and the effective potential
vanishes, $V_{\mathrm{eff}}\left( \alpha \right) =0$. So the atomic states
are delocalized. When $U$ is large, as the pumping strength increases, the
system undergoes a second order phase transition from the \textquotedblleft
normal" phase to a \textquotedblleft delocalized superradiant" phase, see
Fig.~\ref{fig2}(b). As a result the effective potential $V_{\mathrm{eff}%
}\left( \alpha \right) $ is non-zero, and the atomic density is modulated
but still delocalized. For larger pumping strength, the system undergoes a
transition into a \textquotedblleft localized superradiant" phase. In this
phase, the effective potential becomes so large that the atomic wave
function is localized. In the small $U$ regime, these two transitions merge
into one first order transition, where the cavity field and the effective
potential suddenly jump to large values [Fig.~\ref{fig2}(c)]. Note that
these conclusions are also valid for the $N$ boson system, with the same
phase diagram unchanged provided we keep $\lambda /J$, $U/J$ invariant and
scale $\Delta /J\rightarrow N\Delta /J$, $\kappa /J\rightarrow N\kappa /J$.
The appearance of a localized superradiant phase is consistent with a
previous study\cite{CAAM2016}.

\begin{figure}[tbp]
\includegraphics[width=3.2in]
{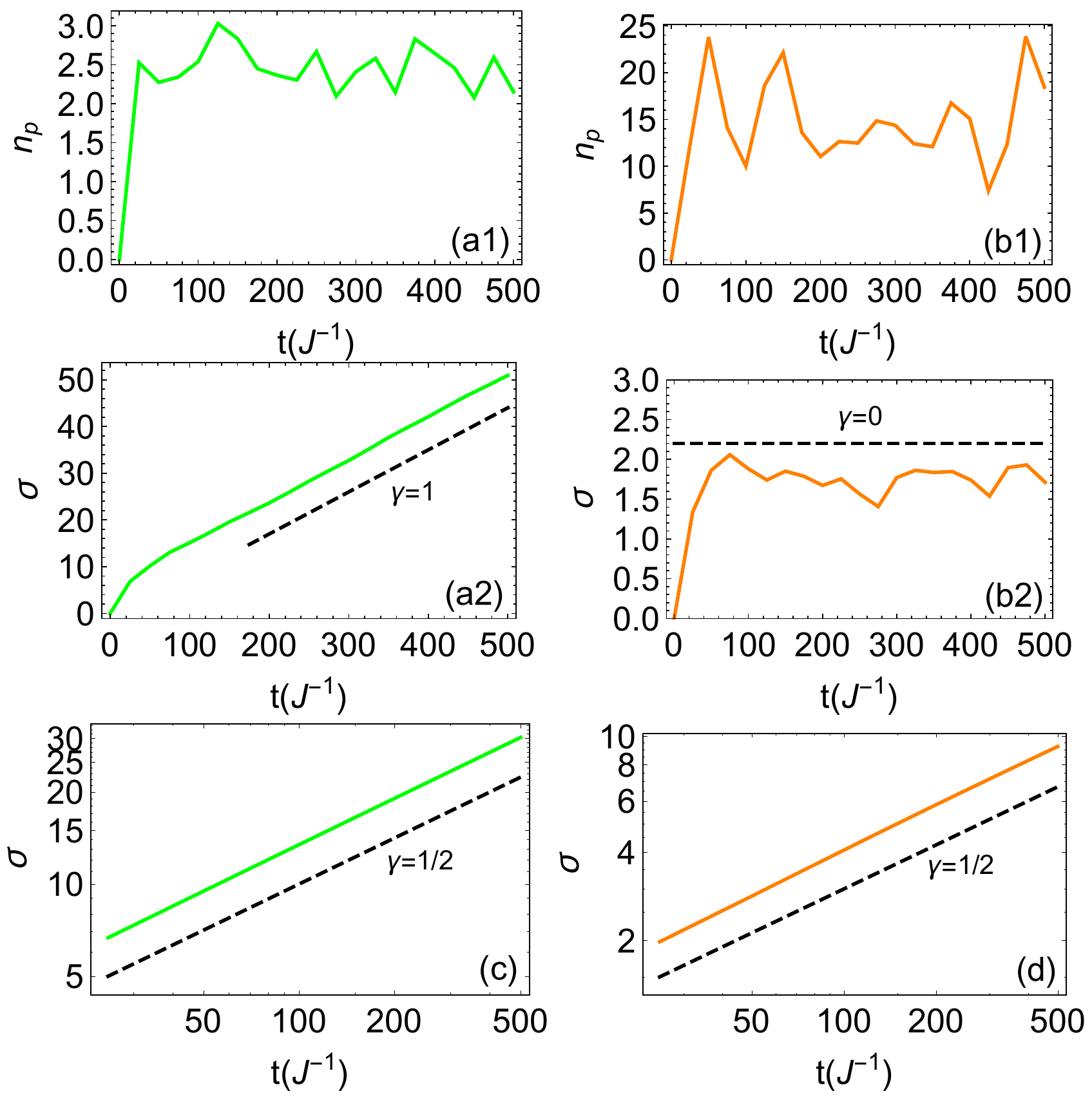}
\caption{(a1), (a2) Dissipationless nonequilibrium dynamics in the
delocalized regime. (a1) The photon number $n_{p}$, (a2) wave packet width $%
\protect\sigma $. The dashed line is a guide for ballistic behaviour $%
\protect\gamma =1$. (b1), (b2) Dissipationless nonequilibrium dynamics in
the localized regime. (c), (d) Dynamics in the large dissipation limit in
delocalized and localized phases of the mean-field model respectively. The
dashed line is a guide for diffusion $\protect\gamma =1/2$. The parameters
of (a1), (a2) and (c) are identical to the dots $d_{1}$ in Fig.~2(a), while
the parameters of (b1), (b2) and (d) are given by the dots $d_{4}$ in
Fig.~2(a).}
\label{fig3}
\end{figure}

\textit{Wavepacket spreading}. Given the fact that the cavity can drive a
transition between extended and localized states, at least at mean-field
level, it is natural to ask how this affects the particle \textit{transport}%
. To answer that, we have investigated how an atomic wave packet spreads in
the driven and damped cavity field. We set the initial state to be the atom
located in the centre of the lattice and the cavity empty. We then consider
turning on the pump laser, and calculate the time evolution of the wave
packet width $\sigma (t)=\sqrt{\left\langle X^{2}\right\rangle -\left\langle
X\right\rangle ^{2}}$, where $X=\sum_{j}j\hat{c}_{j}^{\dag }\hat{c}_{j}$ is
the centre-of-mass of the wave packet. We find, quite generally, that the
width grows as a power-law, $\sigma (t)\sim t^{\gamma }$ at long times.
However, the nature of this growth is a surprisingly subtle issue: its
qualitative form requires an accurate description of the quantum
fluctuations of the driven-damped cavity field. We illustrate this by first
presenting results for two limiting cases of the cavity damping (for these
cases we set $U=0$ for simplicity).

\textit{Dissipationless limit.} We first consider the dissipationless limit, $%
\kappa =0$. The system is then closed and the dynamics is given by
unitary evolution under the Hamiltonian (\ref{Ham}). We numerically simulate
the unitary evolution process, obtaining the photon number $%
n_{p}(t)=\left\langle \hat{a}^{\dag }\hat{a}\right\rangle $ and the wave
packet width $\sigma (t)$, see Fig.~\ref{fig3}. The photon number first rises from zero to a nonzero value in a short time, and then shows large fluctuations around this nonzero value. For small
pumping strength, the average photon number is small and the
wave packet spreading is ballistic, $\gamma =1$ [Fig.~\ref{fig3}(a1)(a2)].
While for large pumping strength, the photon number is larger and the width
finally saturates at long times, indicating localized behaviour, $\gamma =0$
[Fig.~\ref{fig3}(b1)(b2)]. These qualitative forms of dynamics are
consistent with the mean field steady states: at small coupling the steady
state wave function is delocalized; while at large couplings, the steady
state wave function is localized.

\textit{Large dissipation limit. }We now consider the opposite limit, in
which the dissipation $\kappa $ is so large that the lifetime of the cavity
is negligible. In this case, the cavity field will adiabatically follow the
distribution of the atom density, with $\hat{a}\approx -\frac{\lambda }{%
\Delta -i\kappa }\hat{K}$, where $\hat{K}=\sum_{j}u_{j}\hat{c}_{j}^{\dag }%
\hat{c}_{j}$. Since the cavity field is fixed by the atomic density, one can
substitute this formula into the Hamiltonian (\ref{Ham}) and the quantum
master equation (\ref{QME}) to obtain the effective master equation for the
atom as $\partial _{t}\rho _\mathrm{a}=-i\left[ H_{\mathrm{eff}},\rho _\mathrm{a}\right]
+\kappa ^{\prime }\left( 2\hat{K}\rho _\mathrm{a}\hat{K}-\hat{K}^{2}\rho _\mathrm{a}-\rho
_\mathrm{a}\hat{K}^{2}\right) $. Here $\rho _\mathrm{a}$ is the reduced density matrix of
the atoms, and the effective Hamiltonian is $H_{\mathrm{eff}%
}=-J\sum_{j=1}^{L}\left( \hat{c}_{j+1}^{\dag }\hat{c}_{j}+h.c.\right)
+V^{\prime }\sum_{j=1}^{L}u_{j}^{2}\hat{c}_{j}^{\dag }\hat{c}_{j}$, with $%
V^{\prime }=-\frac{2\lambda ^{2}\Delta }{\Delta ^{2}+\kappa ^{2}}$ and $%
\kappa ^{\prime }=\frac{\lambda ^{2}\kappa }{\Delta ^{2}+\kappa ^{2}}$. This
effective model describes an atom hopping in a quasi-periodic lattice with a
global noise, which is imposed by the damped cavity field. We have
numerically solved this effective quantum master equation. The temporal
dynamics of the width of the atomic wave packet is shown Fig.~3(b). We find
that, in this large dissipation limit, the wavepacket always spreads
diffusively $\sigma \sim t^{1/2}$, both where the mean-field solution shows
delocalized [Fig.~\ref{fig3}(c)] and localized [Fig.~\ref{fig3}(d)]
behaviours. Thus, this global noise destroys the coherence and
makes the atom diffuse like a classical Brownian particle at long times.

\begin{figure}[tbp]
\includegraphics[width=3.1in]{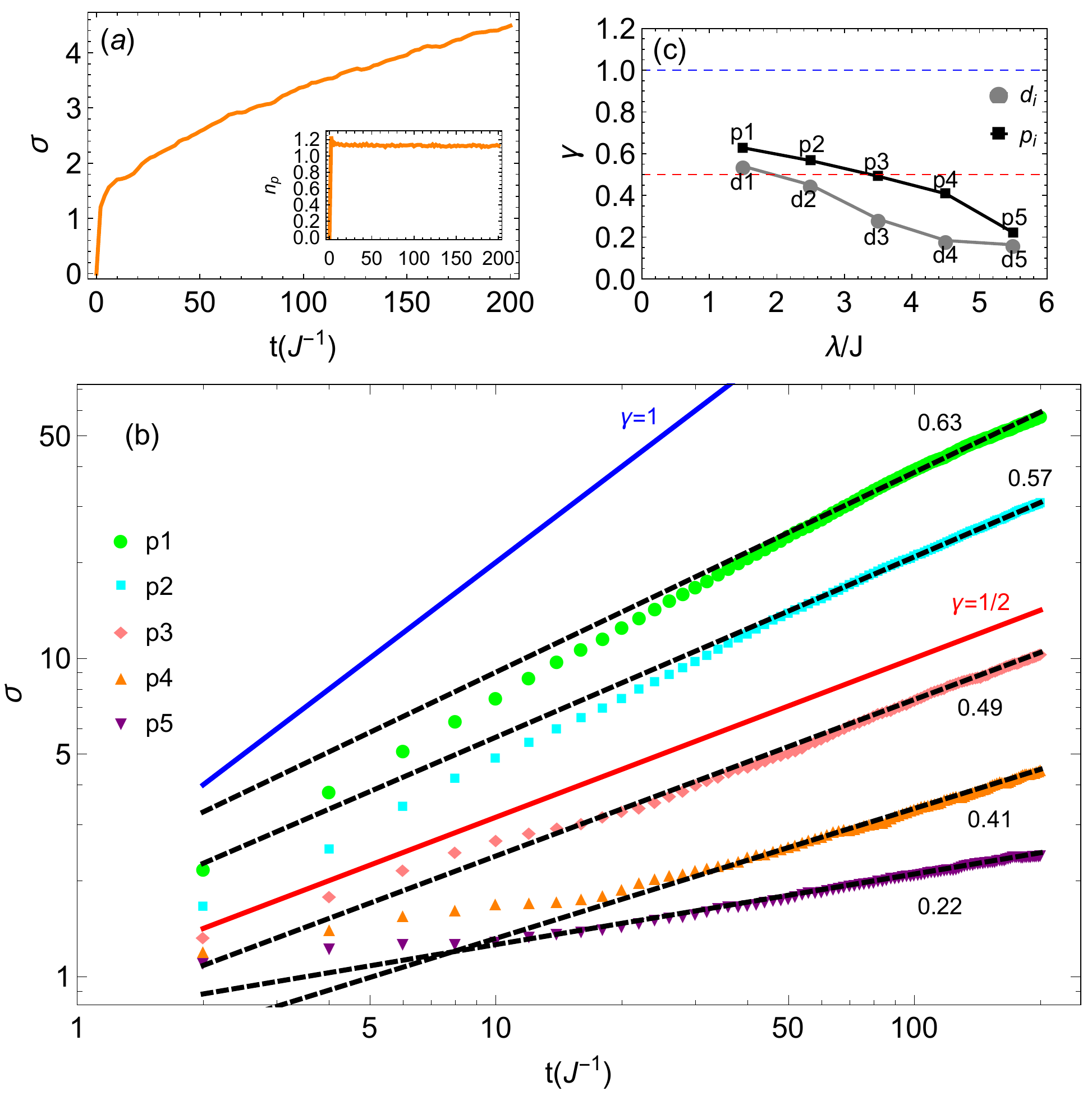}
\caption{(a) Photon number and wave packet width evolution from the exact
quantum trajectory method. The parameters are given by the point $p_{4}$ in
Fig.~2(a). (b) Evolution of the wave packet width. Dashed lines are the
results of fitting to $\protect\sigma =at^{\protect\gamma }$, and the
corresponding numbers are the exponents. The two solid lines represent the
ballistic, $\protect\gamma =1,$ and diffusive, $\protect\gamma =1/2,$ cases.
(c) The exponent $\protect\gamma $ crosses over from super-diffusion to
sub-diffusion with increasing pump strength. The two dashed lines show the
ballistic and diffusive values. The parameters of $d_{i}$ and $p_{i}$ can be
found in Fig.~2(a). }
\label{fig4}
\end{figure}

\textit{Quantum trajectory method. }After considering these two limiting
cases, we now investigate the generic situation, in which the cavity
dissipation is finite. We employ the so-called quantum trajectory method\cite%
{Daley2014}, which is a stochastic way to simulate the quantum master
equation % This
%method simulates both the unitary\ evolution and non-unitary quantum
%jump of the wave function of the system instead of the density matrix
%in a single quantum trajectory. B
by averaging over many quantum trajectories.
%, one can approach to the correct observable.

We have used this method to simulate the wave packet spreading for different
pumping strengths. The results are plotted in Fig.~\ref{fig4}. In Fig.~\ref%
{fig4}(a), one can see a typical dynamics of the system. Similar to the
dissipationless limit, the photon number rises to a nonzero value in a short
time scale. After that, the cavity field enters into a quasi-steady state,
in which the photon number has small fluctuations around its mean. Note that
the fluctuation amplitude is much smaller than that in the dissipationless
limit [Fig.~\ref{fig3}(a1)], as the existence of the cavity dissipation
suppresses these fluctuations. At short times, the width of the wave packet
grows quickly from zero, since the cavity field has not yet built up a large
effective potential. However, at long times, after the cavity field has
reached its quasi-steady state, we find that the wave packet spreads
according to anomalous diffusion, $\sigma \sim t^{\gamma }$, with $0<\gamma
<1$. We find that the exponent $\gamma $ depends on the pumping strength and
other parameters. As shown in Fig.~4(b), when the pumping strength is small,
$\gamma $ is relatively large, corresponding to super-diffusion, $\gamma
>1/2 $. When the pumping strength is large, $\gamma $ becomes relatively
small, crossing over to the sub-diffusion regime, $\gamma <1/2$ [see
Fig.~4(c)]. This behaviour is very different from the dissipationless and
large dissipation limits. This indicates that the observed anomalous
diffusion is a result of both the dissipation and the cavity
dynamics.

\begin{figure}[tbp]
\includegraphics[width=3.2in]
{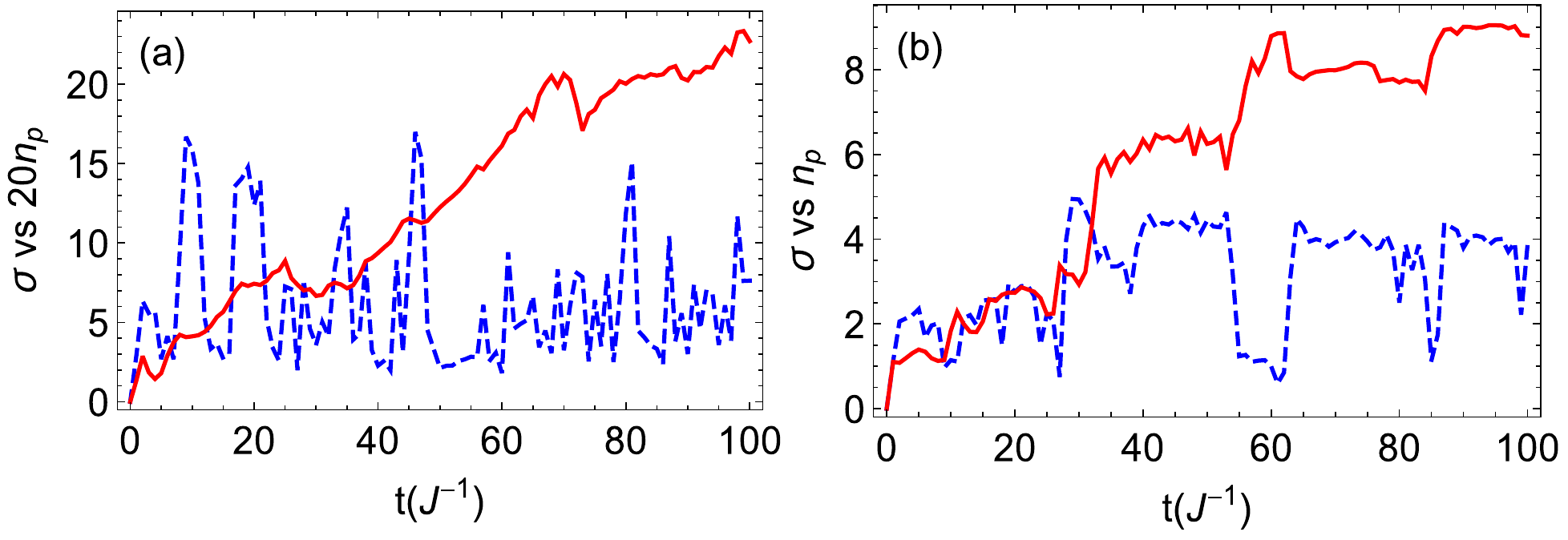}
\caption{Individual quantum trajectories of the stochastic evolution in: (a)
the super-diffusive regime; and (b) the sub-diffusive regime. The solid line
is the wave packet width, and the dashed line is the photon number
[multiplied by 20 in (a)]. The parameters of (a) and (b) are given by the
dots $d_{1}$ and $d_{3}$ in Fig.~2(a).}
\label{fig5}
\end{figure}

How can we understand this anomalous diffusion? We plot the evolution of the
photon number and the wave packet width for a single quantum trajectory in Fig.~\ref{fig5}. Comparing the photon number and the width,
one finds that when the photon number is large the width almost does not
grow: at these times, the effective potential induced by the cavity is very
strong, such that the wave packet is localized and cannot spread freely.
When the cavity field fluctuates to a small value, it reduces the effective
potential: at these times, the wave packet can spread ballistically until
the revival of the photon number. With the help of this picture, we can map
the particle hopping into a L\'{e}vy walk with rests\cite%
{Levy_walk_Rev2015,Levy_walk_book2011}. When the cavity field is lower than
a threshold, the particle moves ballistically at a certain maximal velocity
set by the bandwidth. When the cavity field exceeds the threshold, due to
the quantum localization the motion is switched off, and the particle is at
rest. The time interval of the \textquotedblleft on\textquotedblright and
\textquotedblleft off\textquotedblright , i.e. moving time and waiting time
are random variables, since the cavity is affected by the noise from the
environment. Crucially, we find that, in the large pumping regime, the distribution of
waiting times has a broad tail, leading to subdiffusive behaviour. While
in the small pumping regime, the broad tail of the moving time distribution
dominates, and gives superdiffusion. By increasing the pump strength, one
increases the mean cavity field and decreases the switching threshold. This
gradually tunes the distribution of waiting time and the moving time,
resulting in a crossover from subdiffusion to superdiffusion.

\textit{Final remarks.} Anomalous diffusion is predicted in other quantum
systems with specific forms of coloured noise\cite%
{QAD_noise1987,QAD_noise2006,QAD_noise2006R,QAD_noise2017}. In our model
anomalous diffusion arises naturally in a very simple experimental setting
with generic form of damping. The wave packet spreading could be detected by
situ imaging. However, the anomalous properties could also be detected from
the photons leaking from the cavity\cite{Hemmerich2017Bloch} for which we
predict long-tailed distributions of lower and higher cavity occupations. It
will be interesting to consider situations in higher dimensions, or for
larger particle densities in which cavity-mediated interactions will also
play a role.

\acknowledgments{This work was supported by EPSRC Grant Nos. EP/K030094/1 and EP/P009565/1.}

\end{document}